\documentclass[aps,prb,showpacs,amsmath,amssymb]{revtex4}
\usepackage{graphicx}% Include figure files
\usepackage{bm}% bold math

\begin{document}

\title{Introduction to superconductivity in metals without inversion center}

\author{V. P. Mineev$^1$, M.Sigrist$^2$}

\affiliation{
$^{1}$ Commissariat \`a l'Energie Atomique, INAC/SPSMS, 38054
Grenoble, France\\
$^{2}$ Theoretische Physik ETH-Hongerberg, CH-8093 Zurich, Switzerland
}
\date{\today}

\begin{abstract} This Chapter gives a brief introduction to some basic aspects metals and superconductors in crystal without inversion symmetry. In a first part we analyze some normal state properties which arise through antisymmetric spin-orbit coupling existing in non-centrosymmetric materials and show its influence on the de Haas-van Alphen effect. For the superconducting phase we introduce a multi-band formulation which naturally arises due the spin splitting of the bands by spin-orbit coupling. It will then be shown how the states can be symmetry classified and their relation to the original classification in even-parity spin-singlet and odd-parity spin-triplet pairing states. The general  Ginzburg-Landau functional will be derived and applied to the nucleation of superconductivity in a magnetic field. It will be shown that magneto-electric effects can modify the standard paramagnetic limiting behavior drastically. 
\end{abstract}

%\pacs{}

\maketitle

\section{Introduction}
Motivated by the discovery of  the non-centrosymmetric heavy Fermion superconductor CePt$_3$Si \cite{Bauer04}, the physics of unconventional superconductivity in
materials without inversion symmetry has recently become a subject of growing interest.
The lack of inversion symmetry, a key symmetry for Cooper pairing, combined with unconventional pairing symmetry is responsible for a number of intriguing novel properties. In a short time the list of new superconductors in this class has been enlarged by compounds such as  UIr \cite{Akazawa04}, CeRhSi$_3$ \cite{Kimura05}, CeIrSi$_3$  \cite{Sugitani06}, Y$_2$C$_3$  \cite{Amano04} and Li$_2$(Pd$_{1-x}$Pt$_x$)$_3$B  \cite{LiPt-PdB}. In all listed heavy fermion compounds superconductivity appears in combination with a magnetic quantum phase transition suggesting the presence of strong electron correlation effects. Thus, it is widely believed that magnetic fluctuations are likely responsible for inducing here unconventional  Cooper pairing. For other materials
correlation effects seem to be less relevant. Nevertheless, some of them show unexpectedly features of unconventional pairing. 
While in some cases experimental results give rather clear suggestions on the gap symmetry, the definite identification of pairing states is far from concluded. 

The microscopic theory of superconductivity in metals without inversion has a long history predating these recent experimental 
developments  \cite{BGR76,LNE85,Edel89,GR01}. Specific aspects such as the possibilities  of inhomogeneous superconducting states \cite{Min93,MS94} and the magneto-electric effect \cite{Edel95,Edel96} in this type of materials have been discussed already in the nineties of the last century. Moreover, 
general symmetry aspects of non-centrosymmetric superconductors have been addressed rather early in Refs. \cite{SZB04,SC04,Min04}. A wide variety of physical phenomena connected with non-centrosymmetricity have since been studied by many groups:

\begin{itemize}
\item  paramagnetic limitations of superconductivity and the helical vortex state \cite{Y02,BG02,DF03,Edel03,Sam04,FAKS04,Min05,KAS05,Fuj05,OIM06,AK07}; 

\item paramagnetic susceptibility \cite{FAS04,Sam05,Sam07,YS07} and the magnetic field induced superconducting gap 
structure \cite{Fuj07};

\item Josephson and quasiparticle tunneling  \cite{YTI05,BS06}, surface bound states \cite{IHSYMTS07,VVE08}, and vortex bound states  \cite{LuY08};

\item London penetration depth \cite{HWFS06} and the magnetic field distribution \cite{LY08};

\item effects of impurities \cite{Edel05,FAMS06,MinSam07};

\item upper critical field \cite{Sam08,Sam08'};

\item nuclear magnetic relaxation rate \cite{Sam05',HWFS06'};

\item general forms of pairing interaction \cite{SamMin08};

\item inhomogeneous superconducting states in the absence of external field \cite{MinSam08}.

\end{itemize}

In this Chapter we give a brief introduction to several topics in this context, leaving most of the special aspects of non-centrosymmetric superconductivity to other Chapters.

\section{Normal state}

The absence of inversion symmetry is imprinted into the electronic structure through spin-orbit coupling effects. 
Already the normal state of non-centrosymmetric metals bears intriguing features which result from the specific form of 
spin-orbit coupling. In this section we discuss the electronic spectrum.  
%and the consequences for the de Haas-van Alphen experiments. 

\subsection{Electronic states in non-centrosymmetric metals}
\label{sec: Basics}

Our starting point is the following Hamiltonian of non-interacting
electrons in a crystal without inversion center:
\begin{equation}
\label{H_0}
    H_0=\sum\limits_{{\bf k}}\sum_{\alpha\beta=\uparrow,\downarrow}
[\xi({\bf k})\delta_{\alpha\beta}+\mbox{\boldmath$\gamma$}({\bf k}) 
   \cdot \mbox{\boldmath$\sigma$}
 _{\alpha\beta}]
    a^\dagger_{{\bf k}\alpha}a_{{\bf k}\beta}
\end{equation}
where $ a^\dagger_{{\bf k}\alpha} $ ($a_{{\bf k}\alpha}$) creates (annihilates) an electronic state $ | {\bf k} \alpha \rangle $.  Furthermore, $\xi({\bf k})=\varepsilon({\bf k})-\mu$ denotes the spin-independent part of the spectrum measured relative to the chemical potential $ \mu$, $\alpha,\beta=\uparrow,\downarrow$ are spin indices and $ \mbox{\boldmath$\sigma$}$ are the Pauli matrices. The sum over ${\bf k}$
is restricted to the first Brillouin zone. The second term in Eq.
(\ref{H_0}) describes the  {\it antisymmetric} spin-orbit (SO) coupling whose form depends on the specific non-centrosymmetric crystal structure  \cite{Dressel55,Rashba60}. 
The pseudovector $\mbox{\boldmath$\gamma$}({\bf k})$  satisfies
$\mbox{\boldmath$\gamma$}(-{\bf k})=-\mbox{\boldmath$\gamma$}({\bf k})$ and 
$g\mbox{\boldmath$\gamma$}(g^{-1} {\bf k})=\mbox{\boldmath$\gamma$}({\bf k})$,
where $g$ is any symmetry operation in the generating point group ${\cal G}$ of
the crystal (see below). The usual symmetric spin-orbit coupling which is present also in centrosymmetric crystals
yields a new spinor basis (pseudospinor) $\alpha,\beta$ in Eq. (\ref{H_0}),  which retains the ordinary spin-1/2 structure with complete SU(2)-symmetry. This is different for the antisymmetric spin-orbit coupling. The effect of the 
antisymmetric spin-orbit coupling is a spin splitting of the band energy with $ {\bf k}$-dependent spin quantization axis which removes the SU(2)-symmetry.

Depending on the purpose it is more convenient to express the Hamiltonian 
(\ref {H_0}) in the initial 2x2 matrix form ({\it spinor representation}) or in its diagonal form ({\it band representation}).  The
energy bands are given by 
\begin{equation}
\xi_{\pm}({\bf k})=\xi({\bf k})\pm |\bm{\gamma}({\bf k})| 
\end{equation}
with the Hamiltonian
\begin{equation}
\label{H_0 band}
    H_0=\sum_{{\bf k}}\sum_{\lambda=\pm}\xi_\lambda({\bf k})c^\dagger_{{\bf k}\lambda}c_{{\bf k}\lambda} ,
\end{equation}
where the two sets of electronic operators are connected by a unitary transformation,
\begin{equation}
a_{{\bf k}\alpha}=\sum_{\lambda}u_{\alpha\lambda}({\bf k})c_{{\bf k}\lambda},
\label{trans}
\end{equation}
with
\begin{equation}
\label{Rashba_spinors}
  ( u_{\uparrow\lambda}({\bf k}),~~ u_{\downarrow\lambda}({\bf k})) =
   \frac{( |\mbox{\boldmath$\gamma$}|+\lambda\gamma_z ,~~ \lambda (\gamma_x+i\gamma_y) )}{\sqrt{2|\mbox{\boldmath$\gamma$}|(|\mbox{\boldmath$\gamma$}|+\lambda\gamma_z)}}.
\end{equation}
 The normal-state electron Green's functions in the spinor representation can be written as
\begin{equation}
\label{G spin}
\hat{G}({\bf k},\omega_n)=
\sum_{\lambda=\pm}\hat\Pi_\lambda({\bf k})G_\lambda({\bf k},\omega_n),
\end{equation}
where
\begin{equation}
\label{Pis}
    \hat\Pi_\lambda({\bf k})=\frac{1+\lambda\hat{\mbox{\boldmath$\gamma$}}({\bf k})
\mbox{\boldmath$\sigma$}}{2}
\end{equation}
are the band projection operators  and $\hat{\mbox{\boldmath${\gamma}$}}=\mbox{\boldmath$\gamma$}/|\mbox{\boldmath$\gamma$}|$. The Green's functions in the
band representation have then the simple form  
\begin{equation}
\label{band GF}
    G_\lambda({\bf k},\omega_n)=\frac{1}{i\omega_n-\xi_\lambda({\bf k})} \; ,
\end{equation}
where $\omega_n=\pi T(2n+1)$ is the Matsubara frequency.

The Fermi surfaces defined
by the equations $\xi_\pm({\bf k})=0$ are split, except at specific points
or lines where $|\mbox{\boldmath$\gamma$}({\bf k})|=0$ is satisfied. The band dispersion functions
$\xi_\lambda({\bf k})$ are invariant with respect to all operations
of ${\cal G}$ and the time reversal operations $ K=i\hat\sigma_2K_0 $ ($K_0$ is the complex conjugation). 
 The states $|{\bf k},\lambda\rangle$ and
$K|{\bf k},\lambda\rangle$ belonging to the band energies $ \xi_{\lambda}({\bf k})$ and $\xi_{\lambda}(-{\bf k})$, respectively, are degenerate, since 
the time reversal operation yields
$K|{\bf k},\lambda\rangle=t_\lambda({\bf k})|-{\bf k},\lambda\rangle$,
where $t_\lambda({\bf k})=-t_\lambda(-{\bf k})$ is a nontrivial phase
factor \cite{GR01,SC04}.  For the eigenstates of $ H_0$, defined by 
(\ref{Rashba_spinors}), this phase factor takes the form,
\begin{eqnarray}
\label{t lambda}
    t_\lambda({\bf k})=-\lambda
    \frac{\gamma_x({\bf k})-i\gamma_y({\bf k})}{\sqrt{\gamma_x^2({\bf k})+\gamma_y^2({\bf k})}}.
\end{eqnarray}

Finally we turn to the basic form of the antisymmetric spin-orbit coupling as it results from 
the non-centrosymmetric crystal structures. Here we ignore the Brillouin zone structure 
and use only the expansion for small momenta $ {\bf k} $ leading to basis functions satisfying
the basic symmetry requirements of $ \bf{\gamma}({\bf k}) $. 
For the cubic group ${\cal G}= \mathbf{O} $,  the point group of
Li$_2$(Pd$_{1-x}$,Pt$_x$)$_3$B, the simplest form compatible with
symmetry requirements is
\begin{equation}
\label{gamma_O}
    \mbox{\boldmath$\gamma$}({\bf k})=\gamma_0{\bf k},
\end{equation}
where $\gamma_0$ is a constant. For point groups containing
improper elements, i.e. reflections and rotation-reflections,
expressions become more complicated. The full
tetrahedral group ${\cal G}=\mathbf{T}_d$, which is relevant for
Y$_2$C$_3$ and possibly KOs$_2$O$_6$, the expansion of $ \mbox{\boldmath$\gamma$} ({\bf k}) $ 
starts with third order in the momentum, 
\begin{equation}
\label{gamma_Td}
    \mbox{\boldmath$\gamma$}({\bf k})=\gamma_0[k_x(k_y^2-k_z^2)\hat x+k_y(k_z^2-k_x^2)\hat
    y+k_z(k_x^2-k_y^2)\hat z].
\end{equation}
This is sometimes called Dresselhaus spin-orbit coupling \cite{Dressel55},
and was originally discussed for bulk
semiconductors of zinc-blend structure. 

The tetragonal point group
${\cal G}=\mathbf{C}_{4v}$, relevant for CePt$_3$Si,
CeRhSi$_3$ and CeIrSi$_3$, yields the antisymmetric spin-orbit coupling 
\begin{equation}
\label{gamma C4v}
\mbox{\boldmath$\gamma$}({\bf k})=\gamma_\perp(k_y\hat x-k_x\hat y)
    +\gamma_\parallel k_xk_yk_z(k_x^2-k_y^2)\hat z.
\end{equation}
In the purely two-dimensional case, setting $\gamma_\parallel=0$
one recovers the Rashba interaction \cite{Rashba60} which is often
used to describe the effects of the absence of mirror symmetry in
semiconductor quantum wells.

\subsection{de Haas - van Alphen effect}

An experimental way of observing the spin-splitting of the Fermi surface is the de Haas - van Alphen effect which can help to estimate the magnitude of the antisymmetric spin-orbit coupling \cite{MinSam05}.
The single-electron Hamiltonian (\ref{H_0}) can be extended to include the magnetic field as follows:
\begin{equation}
 H_0=\sum_{{\bf k}}\sum_{\alpha\beta=\uparrow,\downarrow}
    [\xi({\bf k})\delta_{\alpha\beta}+\mbox{\boldmath$\gamma$}({\bf k})\mbox{\boldmath$\sigma$}
    _{\alpha\beta}-\mu_B{\bf H}\mbox{\boldmath$\sigma$} _{\alpha\beta}]
    a^\dagger_{{\bf k}\alpha}a_{{\bf k}\beta}.
\label{e1}
\end{equation}
The last term describes the Zeeman interaction
for an external magnetic field ${\bf H}$, with $\mu_B$ being the Bohr
magneton. 
%[using a general form of the Zeeman energy for band electrons, $\mu_{ij}({\bf k})H_i\sigma_j$, would not add anything to the substance of our results]
The orbital effect of the field can
be included by replacing ${\bf k}\to{\bf k}+(e/\hbar
c)\bf{A}(\hat{\bf{r}})$,\cite{LL9} where $\hat{\bf{r}}=i\bf{\nabla}_{{\bf k}}$
is the position operator in the ${\bf k}$-representation.
% and $e$ is the absolute value of the electron charge. 
  
%For the tetragonal group $\mathbf{C}_{4v}$, the symmetry of both CePt$_3$Si and LaPt$_3$Si, the simplest form of $\bm{\gamma}({\bf k})$ compatible with the symmetry requirements is given by (\ref{gamma C4v}). One can not expect Eq. (\ref{gamma C4v}) to fully reproduce the spin-orbit splitting in $CePt_3Si$ and $LaPt_3Si$ which have quite complicated, multi-sheet Fermi surfaces. Nevertheless, this expression already captures of the spin-orbit coupling, including the qualitative difference in the ${\bf k}$ dependencies of $\gamma_{x,y}({\bf k})$ and $\gamma_{z}({\bf k})$, the presence of a band degeneracy line at $k_x=k_y=0$, and the vanishing of $\gamma_z({\bf k})$ in the high symmetry planes. A natural question is whether one can determine the strength both the $xy$ and $z$ components of spin-orbit coupling using dHvA experiment.
  
The eigenvalues of the Hamiltonian (\ref{e1}) are
\begin{equation}
    \xi_{\lambda}({\bf k},{\bf H})=\xi({\bf k})+\lambda
    |\mbox{\boldmath$\gamma$}({\bf k})-\mu_B{\bf H}|.
\label{e3}
\end{equation}
There are two Fermi surfaces determined by the equations
\begin{equation}
\label{e4}
    \xi_{\lambda}({\bf k},{\bf H})=0.
\end{equation}
For certain directions and magnitudes of $ \bf{H} $ there may be accidental degeneracies of the
Fermi surfaces, determined by the equation $\mbox{\boldmath$\gamma$}({\bf k})=\mu_B{\bf H}$. However, there
are no symmetry reasons for such intersections.
%These three equations can have solutions at some isolated points in the first Brillouin zone, which may or may not be on the Fermi surface.

An important property of the Fermi surfaces (\ref{e4}) is the fact that 
their shapes depend on the magnetic field in a characteristic way, 
which can be directly probed by dHvA experiments. Note that while at $H=0$
time reversal symmetry guarantees $\xi_{\lambda}(-{\bf k})=\xi_{\lambda}({\bf k})$, 
the loss of time reversal symmetry for $ H \neq 0 $ yields, in general, 
$\xi_{\lambda}(-{\bf k},{\bf H})\ne\xi_{\lambda }({\bf k},{\bf H})$, i.e. the
Fermi surfaces do not have inversion symmetry.

Including now the coupling of the magnetic field to the orbital motion of the
electrons we derive in quasi-classical approximation the Lifshitz-Onsager 
quantization rules \cite{LL9} :
\begin{equation}
    S_{\lambda }(\epsilon, k_H)=\frac{2\pi eH}{\hbar c}
    \left[n+\alpha_\lambda(\Gamma)\right]. \label{e5}
\end{equation}
Here $S_\lambda$ is the area of the quasi-classical orbit
$\Gamma$, in the ${\bf k}$-space defined by the intersection of the
constant-energy surface $\epsilon_\lambda({\bf k})=\epsilon$ with the
plane ${\bf k}\cdot\hat{\bf h}=k_H$ ($\hat{\bf h}={\bf H}/H$). Moreover, $n$ is an
integer number ($n\gg 1$), and $0\leq\alpha_\lambda(\Gamma)<1$ 
is connected with the Berry phase of the electron as it moves along 
$\Gamma$ \cite{MS99,Hald04}.
The value of $\alpha_\lambda(\Gamma)$ does not affect the dHvA frequency
discussed below.

The dHvA signal contains contributions from both bands and can be
approximately decomposed into the form,
\begin{equation}
    M_{osc}=\sum_\lambda A_\lambda\cos\left(\frac{2\pi F_\lambda}{H}
    +\phi_\lambda\right), \label{e6}
\end{equation}
where $A_\lambda$ and $\phi_\lambda$ are the amplitudes and the
phases of the oscillations. The amplitudes are given by the standard Lifshitz-Kosevich 
formula and the dHvA frequencies $F_\lambda$ are
related to the extremal (with respect to $k_H$) cross-section
areas of the two Fermi surfaces, 
\begin{equation}
    F_\lambda=\frac{\hbar c}{2\pi e}S_\lambda^{ext}. \label{e7}
\end{equation}
 In addition to the fundamental harmonics (\ref{e6}), the observed dHvA signal also contains 
 higher harmonics with frequencies given by multiple integers of $F_\lambda$.

It is interesting to consider the field dependence of the  band energies (\ref{e3}), 
which yield 
 \begin{equation}
   S_\lambda^{ext}({\bf H})=
    S_\lambda^{ext} (0) + A_\lambda(\hat{\bf h})  H +
    B_\lambda(\hat{\bf h}) H^{2} + \ldots \; .
\label{e8}
\end{equation}
Inserting this in Eq.(\ref{e6}) the term linear in $H$ contributes to the phase shift, 
similar to the paramagnetic splitting of Fermi surfaces in centrosymmetric metals.
The quadratic term is responsible of the magnetic field dependence of the dHvA 
frequencies. This is a specific feature of non-centrosymmetric metals which 
could be observable, if the Zeeman energy is at most of comparable magnitude as
the spin-orbit coupling.

For illustration, let us look at the 
example of a three-dimensional elliptic Fermi surface with
$
    \xi({\bf k})=\frac{\hbar^2k_\perp^2}{2m_\perp}+
    \frac{\hbar^2k_z^2}{2m_z}
    -\epsilon_F,
$
where $m_\perp$ and $m_z$ are the effective masses.
The extremal (maximum) cross-sections of the
Fermi surfaces (\ref{e4}) correspond to $k_z=0$.  Introducing
the Fermi wave vector $k_F$ via
$\epsilon_F=\hbar^2k_F^2/2m_\perp$, we obtain
\begin{equation}
    S_\lambda^{ext}({\bf H})=\pi k_F^2\left[1-\lambda\frac{|\gamma_\perp|k_F}{\epsilon_F}
    \left(1+\frac{\mu_B^2H^2}{2\gamma_\perp^2k_F^2}\right)\right].
\label{example2}
\end{equation}
In this approximation we assumed that  the Zeeman energy is
small compared to the spin-orbit band splitting, which in turn is
much smaller than the Fermi energy: $\mu_BH\ll|\gamma_\perp|
k_F\ll\epsilon_F$. Based on this result it is also possible to obtain an
estimate of the strength of the spin-orbit coupling.

We use the expressions (\ref{e7}) and (\ref{example2}) to calculate
the difference of the dHvA frequencies for the split bands:
\begin{equation}
    F_--F_+=\frac{2c}{\hbar e}|\gamma_\perp |k_Fm_\perp
    \left(1+\frac{\mu_B^2H^2}{2\gamma_\perp^2k_F^2}\right). \label{e9}
\end{equation}
For example, from the frequencies of the ''$\alpha$'' and ''$\beta$''
dHvA frequency branches in LaPt$_3$Si \cite{Hash04},
$F_{\alpha}=1.10\times 10^8$Oe and $F_{\beta}=8.41\times 10^7$Oe,
and $m_\perp\simeq 1.5m$, we obtain for the spin-orbit splitting
of the Fermi surfaces: $|\gamma_\perp |k_F\simeq 10^3$K, which is in
reasonable agreement with the results of band structure
calculations \cite{Hash04,SZB04}. According to Eq. (\ref{e9}), the
magnetic field effect on $F_- - F_+$ in the range of fields used in
Ref. \cite{Hash04} (up to $17$T) should be of the order of a
few percent. In this way the interplay of the Zeeman
splitting and the spin-orbit coupling results in a deformation of
the Fermi surface is responsible for a field dependence of
the dHvA frequencies, an effect absent in centrosymmetric metals.

\section{Superconducting state}

In this section we turn to the discussion of some novel aspects of the superconducting state 
in non-centrosymmetric materials. Here we can consider only a few examples, while a wider
range of other phenomena will be discussed in other Chapters of this book.

\subsection{Basic equations}

After our introductory discussion of single-electron properties we now include electron-electron interactions to examine the implications of non-centrosymmetricity on Cooper pairing. Therefore we retain among all interactions only those terms 
corresponding to the Cooper channel and formulate it in the band representation. The general form is given by
\begin{equation}
\label{H int band gen}
    H_{int}=\frac{1}{2{\cal V}}\sum_{{\bf k}, {\bf k}'}\sum_{\lambda_{1}\lambda_2\lambda_3\lambda_4}
    V_{\lambda_1\lambda_2\lambda_3\lambda_4}({\bf k},{\bf k}')
     c^\dagger_{{\bf k},\lambda_1}
    c^\dagger_{-{\bf k},\lambda_2}c_{-{\bf k}',\lambda_3}
    c_{{\bf k}' ,\lambda_4}.
\end{equation}
It is reasonable to assume  $\lambda_1=\lambda_2$ and $\lambda_3=\lambda_4$, such that
only intra-band pairing is considered. Inter-band pairing is usually suppressed, 
since the spin-orbit coupling induced  band splitting would require that electrons far from the 
Fermi surfaces would have to pair which is unlikely, if the energy scale of the band splitting strongly 
exceeds the superconducting energy scale \cite{footnote}. 
Introducing the notation $\lambda_1=\lambda_2=\lambda$ and
$\lambda_3=\lambda_4=\lambda'$, we obtain:
\begin{equation}
\label{H int reduced}
    H_{int}=\frac{1}{2{\cal V}}\sum\limits_{{\bf k}{\bf k}'\bf{q}}\sum_{\lambda\lambda'}
  V_{\lambda\lambda'}({\bf k},{\bf k}')
    c^\dagger_{{\bf k},\lambda}
    c^\dagger_{-{\bf k},\lambda}c_{-{\bf k}',\lambda'}c_{{\bf k}',\lambda'},
\end{equation}
where 
\begin{equation}
\label{H int reduced'}
V_{\lambda \lambda'}({\bf k},{\bf k}')=t_\lambda({\bf k})t^*_{\lambda'}({\bf k}')
\tilde V_{\lambda\lambda'}({\bf k},{\bf k}') .
\end{equation}
Since under time reversal the creation and annihilation operators behave as
\begin{equation}
Kc^{\dag}_{ {\bf k}, \lambda}=t_{\lambda}¥({\bf k})c^{\dag}_{
-{\bf k}, \lambda}, ~~~~~~~~~Kc_{ {\bf k},
\lambda}=t_{\lambda}¥^{*}¥({\bf k})c_{ -{\bf k}, \lambda},
\label{e82}
\end{equation}
$\tilde V_{\lambda\lambda'}({\bf k},{\bf k}')$ represents the
pairing interaction between time-reversed states. 
The amplitude $\tilde V_{\lambda\lambda'}({\bf k},{\bf k}')$ 
is even in both $\bf k$ and $\bf k'$ due to the
anticommutation of fermionic operators
and is invariant under
the point group operations: $\tilde
V_{\lambda\lambda'}(g{\bf k},g{\bf k}')=\tilde
V_{\lambda\lambda'}({\bf k},{\bf k}')$.
The gap functions of the superconducting state can be
expressed as,
$\Delta_\lambda({\bf k})=t_\lambda({\bf k})\tilde\Delta_\lambda({\bf k})$,
in each band, where $\tilde\Delta_\lambda$ transforms according to one of the
irreducible representations of the crystal point group \cite{Book,SU91}. 

The {\it Gor'kov equations} in each band read
\begin{eqnarray}
&&\left(i\omega_{n}-\xi_{\lambda} ({\bf k}) \right) G_{\lambda}({\bf
k},\omega_{n})+\tilde  \Delta_{{\bf k}\lambda} F_{\lambda}^{\dagger}({\bf
k},\omega_{n})=1  \\
&&\left(i\omega_{n}+\xi_{\lambda} (-{\bf k})\right) F_{\lambda}^{\dagger}¥ ({\bf
k},\omega_{n})+\tilde\Delta_{{\bf k}\lambda}^{\dagger}¥  G_{\lambda}({\bf
k},\omega_{n})=0.
\label{90}
\end{eqnarray}
The gap functions obey the self-consistency equations 
\begin{equation}
\tilde\Delta_{{\bf k}\lambda}=-T\sum_{n}\sum_{{\bf
k}'}\sum_{\lambda'} \tilde V_{\lambda\lambda'}\left( {\bf k},{\bf k}'\right)
F_{\lambda'}({\bf k}',\omega_{n}).
\label{gap eq}
\end{equation}
The resulting Green's functions are then, 
\begin{eqnarray}
G_{\lambda}\left({\bf k},\omega_{n}\right) &=&
\frac{i\omega_{n}+\xi_{\lambda}(-{\bf k})}
{(i\omega_{n}-\xi_{\lambda}({\bf k}))(i\omega_{n}+\xi_{\lambda}(-{\bf k}))
-\tilde{\Delta}_{{\bf k}\lambda} \tilde{\Delta}_{{\bf k}\lambda}^{\dagger}¥ } \\
F_{\lambda}\left({\bf k},\omega_{n}\right)&=& \frac{-\tilde \Delta_{{\bf k}\lambda} 
}{(i\omega_{n}-\xi_{\lambda}({\bf k}))(i\omega_{n}+\xi_{\lambda}(-{\bf k}))
-\tilde{\Delta}_{{\bf k}\lambda}\tilde{\Delta}_{{\bf k}\lambda}^{\dagger}¥ }.
\label{e92}
\end{eqnarray}
and the quasiparticle excitation energies for each band have the form
\begin{equation}
E_{{\bf k}\lambda}¥=
\frac{\xi_{\lambda}({\bf k})-\xi_{\lambda}(-{\bf k})}{2}\pm\sqrt
{\left(\frac{\xi_{\lambda}({\bf k})+\xi_{\lambda}(-{\bf k})¥}{2}
\right)^{2}¥+
\tilde \Delta_{{\bf k}\lambda}\tilde \Delta_{{\bf k}\lambda}^{\dagger}¥ }
\label{e93}
\end{equation}
which becomes in case of time reversal symmetry,
\begin{equation}
E_{{\bf k}\lambda} = \pm \sqrt{\xi_{ \lambda}({\bf k})^2 + \tilde \Delta_{{\bf k}\lambda} \tilde \Delta_{{\bf k}\lambda}^{\dagger}} .
\end{equation}

These forms are analogous to that of a multi-band superconductor  \cite{SMW59}, apart from the fact
that in the non-centrosymmetric case the two bands to not possess spin degeneracy but rather correspond to
a type of spinless fermions, since their spinors on each band are subject to a momentum dependent projection. 
This distinction becomes more apparent, if we
write the  Gor'kov equations in the initial spinor basis (spin up and down),
\begin{eqnarray}
&&\left(i\omega_{n}-\xi({\bf k}) - \mbox{\boldmath$\gamma$}_{\bf
k}\mbox{\boldmath$\sigma$}\right)\hat G({\bf k},\omega_{n})+ \hat\Delta_{{\bf
k}} \hat F^{\dagger}({\bf k},\omega_{n})=\hat 1 \\
&&\left(i\omega_{n}+\xi(-{\bf k}) +  \mbox{\boldmath$\gamma$}_{-{\bf
k}}¥\mbox{\boldmath$\sigma$}^{t}\right)\hat  F^{\dagger}  ({\bf
k},\omega_{n})+\hat \Delta_{{\bf k}}^{\dagger} \hat G({\bf
k},\omega_{n})=0,
\label{e94}
\end{eqnarray}
where
$\xi({\bf k})¥=\varepsilon_{{\bf k}}¥-\mu$,
\begin{equation}
\hat G({\bf k},\omega_{n})=\hat{\Pi} _{+}¥ G_{+}¥({\bf k},\omega_{n})+
\hat{\Pi}_{-}¥ G_{-}¥({\bf k},\omega_{n}),
\label{e95}
\end{equation}
\begin{equation}
\hat F^{\dagger}({\bf k},\omega_{n})=\hat g^{t}¥\left\{\hat {\Pi}_{+}¥
F^{\dagger}_{+}¥ ({\bf k},\omega_{n})+\hat
{\Pi}_{-}¥F^{\dagger}_{-}¥({\bf k},\omega_{n})\right\},
\label{e96}
\end{equation}
\begin{equation}
\hat \Delta_{{\bf k}}=\left \{\hat {\Pi}_{+} \tilde \Delta_{{\bf k},+}+ {\Pi}_{-}¥\tilde \Delta_{{\bf k},-}\right\}\hat g,
\label{e97}
\end{equation}
with $\hat g=i\hat \sigma_{y} $. Examining the form of the gap function reveals that in the non-centrosymmetric superconductor
both even-parity spin-singlet and odd-parity spin-triplet pairing are mixed, since no symmetry is available to distinguish the two. 
Therefore, we may write
\begin{equation}
\hat \Delta_{{\bf k}}=
\frac{\tilde \Delta_{{\bf k},+}+\tilde \Delta_{{\bf k},-}}{2}\hat g+
\frac{\tilde \Delta_{{\bf k},+}-\tilde \Delta_{{\bf k},-}}{2}
\hat{\mbox{\boldmath$\gamma$}}_{\bf k}\mbox{\boldmath${\sigma}$}\hat g .
\label{e98}
\end{equation}
The odd-parity component is represented by a vector which is oriented 
 along the vector $\mbox{\boldmath$\gamma$}_{\bf k}$ of the spin-orbit coupling. In this discussion the only
approximation entering so far is the absence of inter-band pairing.

\subsection{Critical temperature }

For the discussion of the instability condition at the critical temperature and the topology of the quasiparticle gap
we can use  the symmetry properties of the pairing interaction matrix $\tilde{V}_{\lambda\lambda'}({\bf k},{\bf k}')$ mentioned earlier. 
 The momentum dependence of the matrix elements can be represented 
in a spectral form decomposed in products of the basis functions of irreducible representations of
${\cal G}$. It is generally sufficient to consider only the part of the pairing potential based on
one irreducible representation $\Gamma$  corresponding to the superconducting state with maximal critical temperature \cite{Book}.
In a simplified formulation this could be represented by even basis functions on the two bands, $\phi_{+,i}({\bf k})$ and $\phi_{-,i}({\bf k})$,
\begin{equation}
\label{V expansion}
    \tilde V_{\lambda\lambda'}({\bf k},{\bf k}')=-V_{\lambda\lambda'}
    \sum_{i=1}^{d}\phi_{\lambda,i}({\bf k})\phi_{\lambda',i}^*({\bf k}'),
\end{equation} 
While  $\phi_{+,i}({\bf k})$ and $\phi_{-,i}({\bf k})$ both belong to same symmetry representation, their momentum dependence does not have to be exactly the same. The basis functions are
assumed to satisfy the following orthogonality conditions:
$\langle\phi^*_{\lambda,i}({\bf k})\phi_{\lambda,j}({\bf k})\rangle_\lambda=\delta_{ij}$,
where the angular brackets denote the averaging over the
$\lambda$-th Fermi surface.
 The coupling constants
$V_{\lambda\lambda'}$ form a Hermitian matrix, which becomes
real symmetric, if the basis functions are real. 
The gap functions take the form
\begin{equation}
\label{gaps}
    \tilde\Delta_\lambda({\bf k})=\sum_{i=1}^{d_\Gamma}\eta_{\lambda,i}\phi_{\lambda,i}({\bf k}),
\end{equation}
and $\eta_{\lambda,i}$ are the superconducting order parameter
components in the $\lambda$-th band. 

As an example, consider a superconducting state with the order parameter transforming according to a one-dimensional representation 
$\tilde\Delta_\lambda({\bf k})=\eta_{\lambda}\phi_{\lambda}({\bf k})$. The linearized gap equations Eq. (\ref{gap eq}) acquire simple algebraic form
\begin{eqnarray}
\eta_{+}¥ &=&(g_{++}¥\eta_{+}¥+
g_{+-}¥\eta_{-}¥)S_1(T),\nonumber\\
\eta_{-}¥&=&(g_{-+}¥\eta_{+}¥+
g_{--}¥\eta_{-}¥)S_1(T),
\label{e49}
\end{eqnarray}
where
\begin{equation}
\label{g def}
    g_{\lambda\lambda'}=V_{\lambda\lambda'}N_{0\lambda'},
\end{equation}
and $N_{0\lambda}=\langle|\phi_{\lambda}({\bf k})|^2N_{0\lambda}(\hat{{\bf k}})\rangle_\lambda$ is the weighted average  angular dependent density of states over the $\lambda$-th Fermi surface. Note that
for multidimensional representations (dimensional $ d_{\Gamma} $) due to the crystal point symmetry the values of $N_{0\lambda}=\langle|\phi_{\lambda,i}({\bf k})|^2N_{0\lambda}(\hat{{\bf k}})\rangle_\lambda $ 
are equal for  all components $i=1,..d_\Gamma$ and all components $\eta_{\lambda,1}, ...\eta_{\lambda,d_\Gamma}$ separately satisfy the same system of equations (\ref{e49}). 

The function $S_1(T)$ is 
\begin{equation}
S_1(T)=2\pi T\sum_{n\ge 0}¥\frac{1}{\omega_{n}¥}=\ln
\frac{2\gamma \epsilon_c}{\pi T}~,
\label{e50}
\end{equation}
where $\ln\gamma=0,577\ldots $ is the Euler constant, $\epsilon_c$ is an
energy cutoff for the pairing interaction, which we assume to be the same for both bands.
From  Eq. (\ref{e49}) we obtain then 
the following expression for the critical temperature:
\begin{equation}
T_{c0}=\frac{2\gamma\epsilon_c}{\pi}\exp{\left (-\frac{1}{g}\right )},
\label{T_{c0}}
\end{equation}
where
\begin{equation}
\label{g}
    g=\frac{g_{++}+g_{--}}{2}+\sqrt{\left(\frac{g_{++}-g_{--}}{2}\right)^2+g_{+-}g_{-+}}
\end{equation}
is the effective coupling constant.  
For multidimensional representations the critical temperature is the same for all $d_\Gamma$
components of ${\eta}_{\lambda,i}$ of the order parameter. The particular combination of amplitudes
${\eta}_{\lambda,i}$   in the
superconducting state below $T_c$ is
determined by the nonlinear terms in the free energy or self-consistent equation, which
depend on the symmetry of the dominant pairing
channel.

The solution of Eq. (\ref{e49}) ($\eta_+,~\eta_-$) corresponding to the eigenvalue
$S_1(T_c)$ determines two unequal order parameter components $\tilde\Delta_\lambda({\bf k})=\eta_{\lambda}\phi_{\lambda}({\bf k})$. 
In the spinor representation (\ref{e98}) both singlet and 
triplet parts of the order parameter are present. Pure singlet or pure triplet pairing occurs
only under rather restrictive conditions.  First, the momentum dependence
of the gap function in both bands is the same $\tilde\Delta_\lambda({\bf k})=\eta_{\lambda}\phi({\bf k})$. Second, 
$g_{++}=g_{--}$ and $g_{+-}=g_{-+}$ is realized. Then we obtain two solutions of equations 
Eq. (\ref{e49}) with
\begin{equation}
\eta_+= \eta_-,
\label{s}
\end{equation}
\begin{equation}
\eta_+=-\eta_-.
\label{t}
\end{equation}
The critical temperature of the state (\ref{s}) corresponding to the singlet part  is $T_{c0}^s=(2\epsilon_{c}/\pi)e^{-1/g_s}$, where $g_s=g_{++}+g_{+-}$. The critical temperature of the spin triplet state (\ref{t}) is $T_{c0}^t=(2\epsilon_{c}/\pi)e^{-1/g_t}$, where $g_t=g_{++}-g_{+-}$. If $g_{+-}>0$ then $T_{c0}^s>T_{c0}^t$ and the phase transition occurs to the state
(\ref{s}). While at  $g_{+-}<0$ we see that  $T_{c0}^t>T_{c0}^s$ and the phase transition occurs to the state
(\ref{t}).

\subsection{Zeros in the quasiparticle gap}

On the one hand, the zeros in the gap for elementary excitations are dictated by the symmetry of superconducting state
or its {\it superconducting class} which is a subgroup ${\cal H} $ of the group of symmetry of the normal state ${\cal G}\times {\cal K} \times U(1)$, where ${\cal G}$ is the point group, ${\cal K}$ is the group of time reversal, $U(1)$ is the gauge group. 
The procedure to find symmetry dictated nodes is described in Ref.\cite{Book}.
Let us consider the  possible superconducting states (\ref{gaps}) and their nodes for $CePt_{3}¥Si$ with point group symmetry  $C_{4v}$.
This group has four one-dimensional irreducible
representations, $A_{1}, A_{2}, B_{1}, B_{2}$, and one
two-dimensional representation, $E$.  Examples of even basis functions of these irreducible
representations are 

\vskip 0.5 cm

\begin{tabular}{l|c|c}
$\Gamma \qquad $ &  $ \qquad \qquad \phi_{\Gamma}({\bf k}) \qquad  \qquad  $ & nodes \\
\hline
& & \\
$A_1$ & $ k_{x}^{2}¥+k_{y}^{2}¥+ck_{z}^{2} $ & -- \\
$A_2$ & $ k_{x}k_{y}(k_{x}^{2}¥-k_{y}^{2}¥) $ & $ \quad k_{x}¥=0, ~k_{y}¥=0, ~k_{x}¥=\pm k_{y} \quad $ \\
$B_1$ & $ k_{x}^{2}¥-k_{y}^{2} $ & $ k_{x}¥=\pm k_{y} $ \\
$ B_2$ & $ k_{x}k_{y} $ & $ k_{x}¥=0,~ k_{y}¥=0 $ \\
%$ E $ & $  (k_{z}k_{x},k_{z}k_{y}) $ & $ k_x = k_y=0 $ or $ k_z $ \\
\end{tabular}

\vskip 0.3 cm

For $E$ state the basis functions are $ \phi_{E1} ({\bf k}) = k_x k_z $ and $ \phi_{E2} ({\bf k}) = k_y k_z $ and it leads to the order parameter for the Fermi surfaces $ \lambda $,
\begin{equation}
\label{gapE}
    \tilde\Delta_\lambda({\bf k})=\eta_{\lambda,1}\phi_{\lambda,E1}({\bf k})+\eta_{\lambda,2}\phi_{\lambda,E2}({\bf k}).
\end{equation}
The symmetry of superconducting state and the corresponding node positions depend on the particular choice of 
amplitudes $\eta_{\lambda,i}$. Under weak coupling conditions the combination generating least nodes is most stable, corresponding here to
$ (\eta_{\lambda,1}, \eta_{\lambda, 2}) = \eta_{\lambda} (1, \pm i) $ and is a time reversal symmetry violating phase with a line node on the plane $ k_z = 0 $ and point nodes at $ k_x = k_y =0 $. 

On the other hand,  the fact that their gaps on the two Fermi surfaces are composed of an even and an odd parity part, can also lead to nodes which are not symmetry protected, as discussed in Ref.\cite{FAMS06}.

\subsection{The amplitude of singlet and triplet pairing states}

The coupling constants $V_{\lambda\lambda'}$ we have used in previous considerations
can be expressed through the real physical interactions between the electrons naturally introduced  in the initial spinor basis where BCS type Hamiltonian has the following form \cite{Book}
\begin{equation} \begin{array}{l}
\label{int}
\displaystyle 
    H_{int}=\frac{1}{4{\cal V}}\sum\limits_{{\bf k}{\bf k}'\bf{q}}\sum_{\alpha\beta\gamma\delta}
  [V^g({\bf k},{\bf k}')(i\sigma_2)_{\alpha\beta}(i\sigma_2)^\dagger_{\gamma\delta} \\  \\
  \displaystyle \qquad \qquad \qquad  +
  V^u_{ij}({\bf k},{\bf k}')(i\sigma_i\sigma_2)_{\alpha\beta}(i\sigma_j\sigma_2)^\dagger_{\gamma\delta}]
    a^\dagger_{{\bf k}+\bf{q},\alpha}
    a^\dagger_{-{\bf k},\beta}a_{-{\bf k}',\gamma}a_{{\bf k}'+\bf{q},\delta},
    \end{array}
\end{equation}
here the amplitudes $V^g({\bf k},{\bf k}')$ and $V^u_{ij}({\bf k},{\bf k}')$ are even and odd with respect to their arguments correspondingly.
The unitary transformation (\ref{trans}) transforms the pairing Hamiltonian (\ref{int}) to the band representation
(\ref{H int band gen}). 
If we neglect  inter-band pairing, it is reduced to
(\ref{H int reduced}) and (\ref{H int reduced'}) with the amplitudes given by the following expression\begin{equation}
\label{tilde V}
    \tilde V_{\lambda\lambda'}({\bf k},{\bf k}')=
  \frac{1}{2}  V^g({\bf k},{\bf k}')(\sigma_0+\sigma_{x})_{\lambda\lambda'}+\frac{1}{2} 
    V^u_{ij}({\bf k},{\bf k}')\hat\gamma_i({\bf k})\hat\gamma_j({\bf k}')(\sigma_0-\sigma_{x})_{\lambda\lambda'}.
\end{equation}
The explicit derivation is given in \cite{SamMin08}, where the similar procedure was also 
made for more general interactions mediated by phonons or spin fluctuations.
It can be shown that the pairing given by the amplitude 
$V^g({\bf k},{\bf k}')$ in the initial spinor basis including the simple $s$-wave pairing $V^g({\bf k},{\bf k}')=const$
does not induce any inter-band pairing channel. 

To illustrate the origination of the singlet and triplet pairing channels 
 let us consider a  
superconductor with tetragonal symmetry $C_{4v}$ and  Rashba spin-orbital coupling 
$\mbox{\boldmath$\gamma$}
({\bf k})=\gamma_\perp(\hat z\times{\bf k})$, for a spherical Fermi surface.
We describe the pairing  by the following model, which is
compatible with all symmetry requirements:
\begin{eqnarray}
\label{v model}
    &&V^g({\bf k},{\bf k}')=-V_g,\nonumber\\
    &&V_{ij}^u({\bf k},{\bf k}')=-V_u(\hat{\bf{\gamma}}_i({\bf k})\hat{\bf{\gamma}}_j({\bf k}')),
\end{eqnarray}
where $V_g$ and $V_u$ are constants.  
This type of pairing interaction yields the superconducting state with full symmetry of the tetragonal group $C_{4v}$ 
transforming according to unit representation $A_1$ both in singlet and in triplet channels.

With Eq. (\ref{tilde V}) we arrive at
 the band representation:
\begin{equation}
    \tilde V_{\lambda\lambda'}({\bf k},{\bf k}')=-\frac{1}{2}V_g(\sigma_0+\sigma_{x})_{\lambda\lambda'}
-\frac{1}{2}V_u(\sigma_0-\sigma_{x})_{\lambda\lambda'} ,
\end{equation}
 Thus, this pairing interaction is even simpler than that (eqn.(\ref{V expansion})) considered in the previous subsection.
So, in our model the gap
functions in the two bands Eqs. (\ref{gaps}) are:
$\tilde\Delta_\lambda({\bf k})=\eta_\lambda\varphi_{A_1}({\bf k})$ wth ${\bf k}$ independent functions  $\varphi_{A_1}({\bf k})=1$. The amplitudes $\eta_\lambda$ satisfy the
equations
\begin{equation}
\label{gap eq isotropic}
    \eta_\lambda=\sum_{\lambda'}g_{\lambda\lambda'}\,\pi
    T\sum_n\frac{\eta_{\lambda'}}{\sqrt{\omega_n^2+\eta^2_{\lambda'}}},
\end{equation}
where 
\begin{equation}
g_{\pm \pm} = \frac{V_g + V_u}{2} N_{0\pm} , \qquad g_{\pm \mp} = \frac{V_g - V_u}{2} N_{0\mp} .
\end{equation}
The critical temperature is given by Eq. (\ref{T_{c0}}). 

According to the Eq.(\ref{e98}) the singlet and triplet parts of the  order parameter are determined by  the order parameter amplitudes in different bands.
For the ratio of triplet to singlet amplitude in
vicinity of $T_c$ we find:
\begin{equation}
\label{gap variation}
    r\equiv\frac{\eta_+-\eta_-}{\eta_++\eta-}=
    \frac{2g_{+-}+g_{++}-g_{--}-\sqrt{\cal D}}{
    2g_{+-}+g_{--}-g_{++}+\sqrt{\cal D}},
\end{equation}
where ${\cal D}=(g_{++}-g_{--})^2+4g_{+-}g_{-+}$. 
It is easy to see that for $V_u=0$  the triplet component of the order parameter  vanishes identically:$~~r=0$. On the other hand,  for  $V_g=0$ the singlet component of the order parameter disappears: $~~r^{-1}=0$. Generally the relative weight of singlet and triplet component in the order parameter depends on the ratio of pairing interactions decomposed into even and odd parity channel. 

A simple BCS type of model with
\begin{equation}
V^g({\bf k}, {\bf k}') = - V_g \qquad \mbox{and} \qquad V_{ij}^u({\bf k},{\bf k}') = 0
\end{equation}
yields in the band representation
\begin{equation}
\tilde{V}^{BCS}_{\lambda\lambda'}({\bf k},{\bf k}')=-\frac{1}{2}V_g(\sigma_0+\sigma_{x})_{\lambda\lambda'}.
\label{BCS}
\end{equation}
and gives rise to purely spin singlet pairing within our notion.

\subsection{Ginzburg-Landau formulation}

The Ginzburg-Landau theory is a very efficient tool to discuss a wide variety of phenomena of the superconducting state, in particular, the instability conditions at the critical temperature. 
We will derive the Ginzburg-Landau functional from a microscopic starting point with the aim to address in the following chapter the influence of  the magneto-electric effect on the nucleation of superconductivity in a magnetic field, i.e. the modification of paramagnetic limiting in a 
non-centrosymmetric metal. 

For this purpose we extend the self-consistent equation (\ref{gap eq}) to the case where magnetic fields are present and the superconducting order parameter has a weak spatial dependence,
\begin{equation}
\tilde\Delta_{\lambda}({\bf k},{\bf q})=T\sum_{n}\sum_{{\bf
k}'}\sum_{\nu} \tilde V_{\lambda\nu}\left( {\bf k},{\bf k}'\right)
G_\nu({\bf k}',\omega_n)G_\nu(-{\bf k}'+{\bf q},-\omega_n)\tilde\Delta_{\nu}({\bf k}',{\bf q}) . 
\label{gap' eq}
\end{equation}
Near the critical temperature where one can use the normal metal Green functions  $G_\lambda^0 ({\bf k},\omega_n) $  which yield then the linearized gap equation to examine the instability condition. This equation can be derived from the free energy functional
of the form
\begin{equation} \begin{array}{ll}
F= & \displaystyle \frac{1}{2}\int \frac{d^3{\bf q}}{(2\pi)^3}\left\{\sum_{\lambda\nu}\eta_{\lambda,i}^*({\bf q})
\tilde V_{\lambda\nu}^{-1}\eta_{\nu,i}({\bf q}) \right.\\ & \\
& \displaystyle  \left. -\sum_{\nu}
T\sum_{n}\int \frac{d^3{\bf k}}{(2\pi)^3}
\tilde\Delta_{\nu}^*({\bf k},{\bf q})G_\nu({\bf k},\omega_n)G_\nu(-{\bf k}+{\bf q},-\omega_n)\tilde\Delta_{\nu}({\bf k},{\bf q})\right\}.
\end{array}
\label{gap''eq}
\end{equation}
The corresponding normal metal electron Green function $G_\lambda({\bf k},\omega_n)=(i\omega-\xi_{\lambda}({\bf k},{\bf H}))^{-1}$ in a magnetic field is determined by the electron energies (\ref{e3}),
\begin{equation}
    \xi_{\lambda}({\bf k},{\bf H})=\xi({\bf k})+\lambda
    |\mbox{\boldmath$\gamma$}({\bf k})-\mu_B{\bf H}|
    \approx \xi_\lambda({\bf k})-\bf{m}_\lambda({\bf k}){\bf H}     , 
\label{e3'}
\end{equation}
where $\xi_\lambda({\bf k})=\xi({\bf k})+\lambda|\mbox{\boldmath$\gamma$}({\bf k})|$ and
the second term on the right-hand side is the analog of the Zeeman
interaction for non-degenerate bands \cite{Sam05} with the form:
\begin{equation}
\label{mu_Rashba}
    \bf{m}_\lambda({\bf k})=\lambda\mu_B\hat{\mbox{\boldmath$\gamma$}}({\bf k}),
\end{equation}
which is valid everywhere except for the vicinity of the band
crossing points, where the approximation of independent
non-degenerate bands fails. In the standard centrosymmetric metal the magnetic field splits the Fermi surfaces into majority and minority spin surfaces. Here the spin-splitting is imposed at the outset by the spin-orbit coupling.  The effect of the magnetic field is a non-centric deformation of the band and shape of the Fermi surfaces. As we will see this will influence the superconducting condensate nucleated in a magnetic field. 

The normal electron Green function is then approximated as
\begin{equation}
\label{G}
    G_{\lambda}({\bf k},\omega_n)=
    \frac{1}{i\omega_n-\xi_\lambda({\bf k})+\bf{m}_\lambda({\bf k}){\bf H}}.
\end{equation}
 Since the gap function weakly depends
on energy in the vicinity of the Fermi surface, one can integrate
the products of two Green's functions with respect to
$\xi_\lambda=\xi_\lambda({\bf k})$:
\begin{equation}
    N_{0\lambda}\int d\xi_\lambda G_{\lambda}({\bf k},\omega_n)
    G_{\lambda}(-{\bf k}+{\bf{q}},-\omega_n)
    =\pi N_{0\lambda} L_{\lambda}({\bf k},{\bf q},\omega_n),
\end{equation}
where
\begin{equation}
\label{L def}
    L_{\lambda}({\bf k},{\bf q},\omega_n)=\frac{1}{|\omega_n|
    +i\Omega_{\lambda}({\bf k},{\bf q})\,\mathrm{sign}\,\omega_n}
\end{equation}
depends only on $ \hat{k} $, the direction of ${\bf k}$,
\begin{equation}
\label{Omega}
    \Omega_{\lambda}({\bf k},\bf{q})=\frac{\bf{v}_\lambda({\bf k})\bf{q}}{2}-\bf{m}_\lambda({\bf k}){\bf H},
\end{equation}
with
$\bf{v}_\lambda({\bf k})=\partial\xi_\lambda({\bf k})/\partial{\bf k}$
being the Fermi velocity in the $\lambda$-th band.

The Ginzburg-Landau free energy in usual coordinate representation (as well as the Ginzburg-Landau equations)
can be obtained from the Taylor expansion
of Eqs. (\ref{gap' eq}) and (\ref{gap''eq}) in powers of $\Omega_{\lambda}({\bf k},\bf{q})$, by 
the replacement
\begin{equation}
\label{q to D}
   {\bf q}\to{\bf D}=-i\nabla_{{\bf r}}+2e{\bf A}(\bf{r})
\end{equation}
in the final expressions. In the following we will put $\hbar=c=1$. The special form of  $ \Omega_{\lambda}({\bf k},\bf{q}) $ introduces a novel gradient terms in the free energy of non-centrosymmetric superconductors.  Instead of powers of ${\bf q}$ it contains powers of 
$\Omega_{\lambda}({\bf k},\bf{q})$. This can can lead to the formation of nonuniform superconducting state known as {\it helical phases} and {\it magneto-electric effect} in a magnetic field.

\section{Magneto-electric effect and the upper critical field}

The term ''magneto-electric effect'' in non-centrosymmetric superconductors encompasses several intriguing features. It has been discussed on a phenomenological level 
by introducing additional  linear gradients terms to the Ginzburg-Landau free energy,  so-called Lifshitz invariants, like
\begin{equation}
\eta^*({\bf r})  \tilde{K}_{ij}H_iD_j\eta({\bf r})
\label{Lif}
\end{equation}
Here $\eta({\bf r})$ denotes the order parameter of superconductor, 
${\bf H}$ is magnetic field and ${\bf D}=-i\nabla-2e{\bf A}$ is the gauge-invariant gradient.
First predicted by Levitov, Nazarov and Eliashberg \cite{LNE85},  the magneto-electric effect was studied microscopically by several authors \cite{Edel89,Edel95,Edel96,Y02,Sam04}. 
In this context several observable effects have been predicted: (i) the existence of a helically twisted superconducting order parameter in a magnetic field in two and three dimensional cases 
and spontaneous supercurrents in a 2D geometry \cite{Edel96,Y02,Sam04,DF03,KAS05,AK07} and near the superconductor surface \cite{OIM06} as well as along junctions of two superconductors with opposite directions of polarization \cite{Fuj05}, (ii) the augmentation of the upper critical field oriented
 perpendicular to the direction of the space parity breaking \cite{Sam04,KAS05},
(iii) magnetic interference patterns of the Josephson critical current for a magnetic field applied perpendicular to the junction \cite {KAS05}.  

The presence of Lifshitz invariants (\ref{Lif}), however, can mislead to invalid conclusions and a careful analysis of different contributions to an effect is mandatory. This is indeed true for the influence of the magneto-electric effect on paramagnetic limiting. Moreover, the notion of helical phase has to be 
considered carefully as it may imply wrong pictures. In this section we would like to give insight into this subtleties by discussing the magnetic field dependence of the effective critical temperature in the Ginzburg-Landau framework.

\subsection{One band case}

Before considering the intrinsic multi-band situation due to the spin splitting of the electron band, for simplicity we restrict ourselves to a one-band situation, i.e. we ignore one of the two bands. This band shall be characterized by an isotropic density of states at the Fermi energy,  $N_{0+}(\hat{\bf k})=N_+$. 

The Ginzburg-Landau free energy for this one-band case with a one-component order parameter can be derived from Eqn. (\ref{gap''eq})
\begin{equation}
F=\frac{1}{2}\int \frac{d^3q}{(2\pi)^3}\left\{\frac{2}{V_{++}}-N_{0+}S_1(T)+N_{0+}S_3(T) \langle(\phi^2({\bf k})\Omega({\bf k},{\bf q}))^2
\rangle 
\right\}|\eta({\bf q})|^2,
\label{e1a}
\end{equation}
where we restrict to the second order terms \cite{Sam04,MinSam07}. This is sufficient to analyze the instability conditions. 
%(\ref{gap''eq}).
Here, $\phi({\bf k})$ describes the superconducting state and is an even function belonging to one of one-dimensional representations
of the point group of the crystal, $\Omega$ is given by (\ref{Omega}), $\langle...\rangle$ means the averaging over the Fermi surface,
the function $S_1(T)$ is given by eqn. (\ref{e50}) and
%(\ref{e50}),
\begin{equation}
S_3(T)=\pi  T\sum_n\frac{1}{|\omega_n|^3}=\frac{7\zeta (3) }{4\pi^2T^2}.
\label{S_3}
\end{equation} 
The Ginzburg-Landau free energy functional in real space can be obtained through a Fourier transformation and leads to 
\begin{equation}
F=\int d^3r\left\{\alpha(T-T_{c0})|\eta|^2
+\eta^*\left [ K_1(D_x^2+D_y^2) +K_2D_z^2
+ K_{ij}H_iD_j+Q_{ij}H_iH_j\right ]\eta\right\},
\end{equation}
where $\alpha=N_{0+}/2T_{c0}$, $T_{c0}=(2\gamma\epsilon_0/\pi)\exp{(-2/V_{++}N_{0+})}$,
\begin{equation}
K_1=\frac{N_{0+}S_3}{8}\langle\phi^2({\bf k})v_x^2({\bf k})\rangle,
~~~~~~~
K_2=\frac{N_{0+}S_3}{8}\langle\phi^2({\bf k})v_z^2({\bf k})\rangle,
\label{KK}
\end{equation}
\begin{equation}
K_{ij}=-\frac{\mu_B N_{0+}S_3}{2}\langle\phi^2({\bf k})
\hat{\bf{\gamma}}_i({\bf k})v_j({\bf k})\rangle,
~~~~~~~~
Q_{ij}=\frac{\mu_B^2N_{0+}S_3}{2}\langle\phi^2({\bf k})
\hat{\bf{\gamma}}_i({\bf k})\hat{\bf{\gamma}}_j({\bf k})\rangle.
\label{KQ}
\end{equation}
The term linear in $ \bf{H} $ incorporates the magneto-electric effects while the term quadratic in $ \bf{H} $ describes the
paramagnetic effect. Therefore $ Q_{ij} |\eta|^2 $ is connected with the change of the paramagnetic susceptibility in the superconducting phase compared with the normal state (Pauli) susceptibility. In particular, $ Q_{ij} $ vanishes when there is no change of the paramagnetic susceptibility. These coefficients have to be compared with those of a spin singlet state of a centrosymmetric superconductor,
$ Q^{(0)}_{ij} = \mu_B^2 N_0 S_3/2 $ which, assuming $ N_{0+} = N_0 $, is larger than above $Q_{ij} $ due to the fact that $ 1 = \langle \phi^2({\bf k}) \rangle \geq
\langle\phi^2({\bf k}) \hat{\bf{\gamma}}_i({\bf k})\hat{\bf{\gamma}}_j({\bf k})\rangle $. 

We consider now the two illustrative cases, the point group $C_{4v} $ and $ D_4 $ which are characterized by the
pseudovectors
\begin{equation} \begin{array}{ll}
\mbox{\boldmath$\gamma$}({\bf k}) = \gamma_{\perp} (k_y\hat{x}  -  k_x\hat{y}) + \gamma_{\parallel} k_x k_y k_z (k_x^2 - k_y^2)\hat{z} & \quad \mbox{for} \quad C_{4v} , \\
\mbox{\boldmath$\gamma$}({\bf k})=\gamma_\perp(k_x\hat x+k_y\hat y)+\gamma_\parallel k_z\hat z & \quad \mbox{for}  \quad   D_4 .
\end{array}
\end{equation}
For symmetry arguments and using above expressions we find the following relations for the coefficients,
\begin{equation} \begin{array}{ll}
K_{xy} = - K_{yx} \neq 0 & \quad \mbox{and} \quad K_{ij} =0 \quad \mbox{otherwise}  \\
Q_{xx} = Q_{yy} \neq Q_{zz} > 0 & \quad \mbox{and} \quad Q_{ij} =0 \quad \mbox{otherwise} 
\end{array}
\end{equation}
for $ C_{4v} $ where presumably $ |K_{zz}| \ll |K_{xy}| $ and $ Q_{zz} \ll Q_{xx} $ due to large number of nodes in the $ {\bf k} $-dependence of the $ \gamma_{\parallel} $-part of $ \mbox{\boldmath$\gamma$}({\bf k}) $ , and
\begin{equation} \begin{array}{ll}
K_{xx} =  K_{yy} \neq 0, K_{zz} \neq 0 & \quad \mbox{and} \quad K_{ij} =0 \quad \mbox{otherwise}  \\
Q_{xx} = Q_{yy} \neq Q_{zz} > 0 & \quad \mbox{and} \quad Q_{ij} =0 \quad \mbox{otherwise} 
\end{array}
\end{equation}
for $ D_4 $.  

\subsubsection{Symmetry $C_{4v}$,  ${\bf H}\parallel \hat z$}

In the case of $C_{4v}$ for the field directed parallel to $z$-axis ${\bf H}=H(0,0,1)$ the terms 
linear in gradients and $ {\bf H} $ is absent. The standard solution $\eta=e^{iq_yy}f(x)$ of the GL equation
\begin{equation}
\left\{\alpha(T-T_{c0})
+K_1\left [-\frac{\partial^2}{\partial x^2}+\left (-i\frac{\partial}{\partial y}+2eHx \right )^2\right ] +Q_{zz}H^2\right\}\eta=0,
\end{equation}
is degenerate in respect to $q_y$.
The magnetic field dependence of critical temperature is
\begin{equation}
T_c=T_{c0}-\frac{2eK_1}{ \alpha}H -\frac{Q_{zz}}{\alpha}H^2
\end{equation}
Both  the orbital (linear in $ H $) and paramagnetic (quadratic in $H $) depairing effect are present. Compared to the ordinary spin-singlet case, however, the effect of the paramagnetic limiting is weaker here due to $ Q_{zz} < Q^{(0)}_{zz} $. It is important to note here that no magneto-electric effect comes into play here. 

\subsubsection{Symmetry $D_4$,  ${\bf H}\parallel \hat z$}

The situation is quite different for the uniaxial crystals with point symmetry group $D_4$ (or $D_6$).
The GL equation includes gradient terms in the field direction and acquires the form
\begin{eqnarray}
\left\{\alpha(T-T_{c0})+K_1\left [-\frac{\partial^2}{\partial x^2}+\left (-i\frac{\partial}{\partial y}+2eHx \right )^2\right ] \right.\nonumber\\
\left.+i K_{zz} H \frac{\partial}{\partial z}-K_2\frac{\partial^2}{\partial z^2} + Q_{zz}H^2\right\}\eta=0.
\end{eqnarray}
The solution can be written as 
\begin{equation}
\eta=e^{iq_yy}e^{iq_zz}f(x),
\end {equation}
which remains degenerate with respect to the wavevector $q_y$, but not with respect to $q_z$ which is used to maximize the critical temperature to
\begin{equation}
T_c=T_{c0}- \frac{2eK_1}{ \alpha}H +\left (\frac{K_{zz}^2}{4 K_2}-Q_{zz}\right )\frac{H^2}{\alpha}.
\end {equation}
This corresponds to the finite wavevector
\begin{equation}
q_z=\frac{K_{zz} H}{2K_2}.
\end {equation}
Note that this wave vector could also be absorbed into the vector potential without changing the physically relevant results: 
$ {\bf A} \to {\bf A} + \nabla \chi $ with $ \chi = - q_z z/2e $. 

The simple paramagnetic depairing effect is weakened due to magneto-electric response of the system. Adjusting the nucleation of the superconducting phase to the shifted Fermi surface, as incorporated in the wavevector $q_z$, recovers some of the strength of the nucleating condensate. This is a specific effect of the non-centrosymmetric superconductor and has its conceptional analogue in the FFLO phase for centrosymmetric spin singlet superconductors, where the condensate also nucleates with finite momentum Cooper pairs
in order to optimize the pairing of degenerate quasiparticles on the split Fermi surface. 

\subsubsection{Symmetry $C_{4v}$,  ${\bf H}\perp \hat z$}

Now we turn the  magnetic field into the basal plane ${\bf H}=H(\cos\varphi,\sin\varphi,0)$, and impose a
gauge to have the vector potential ${\bf A}=Hz(\sin\varphi,-\cos\varphi,0)$.  The corresponding  GL equation take the form 
\begin{equation}
\left\{\alpha(T-T_{c0})
+K_1(D_x^2+D_y^2)
-K_2\frac{\partial^2}{\partial z^2}+ K_{xy}(H_xD_y-H_yD_x) +Q_{xx}H^2
\right\}\eta=0,
\label{3D}
\end{equation}
where
\begin{equation}
D_x=-i\frac{\partial}{\partial x}+ 2eH_yz, \qquad D_y=-i\frac{\partial}{\partial y}-2eH_xz.
\end{equation}
Like in ordinary superconductors the solution of this equation have the Abrikosov form
\begin{equation}
\eta({\bf r})=\exp{\left [i ({\bf p}\times {\bf r})_z\right ]}f(z),
\end{equation}
where we write ${\bf p}=p{\bf H}/H$ as a vector parallel to the magnetic field ($ ({\bf p}\times {\bf r})_z $ denoting the $z$-component of the vector  ${\bf p}\times {\bf r} $), and  $f(z)$ satisfies the resulting renormalized harmonic oscillator equation
\begin{equation}
\left\{\alpha(T-T_{c0})
+K_1 (2eH)^2(z-z_0)^2
-K_2\frac{\partial^2}{\partial z^2} +\left (Q_{xx}-\frac{K_{xy}^2}{4K_1}\right )H^2
\right\}f(z)=0,
\end{equation}
with the shifted equilibrium position
\begin{equation}
z_0=(2eH)^{-1}\left (p+\frac{K_{xy}}{2K_1}H  \right).
\end{equation}
Thus, the vector $ {\bf p} $ is absorbed into the shift $ z_0 $ and does not appear anywhere else in the equation. 
%This is a consequence of the chosen gauge. 
Then the corresponding eigenvalue determines the
magnetic field dependence of optimized critical temperature:
\begin{equation}
T_c=T_{c0}-\frac{2e\sqrt{K_1K_2}}{\alpha}H +\left (\frac { K_{xy}^2}{4K_1}-Q_{xx} \right ) \frac{H^2}{\alpha}.
\label{tc-c4v-hperp}
\end{equation}
In the used gauge the eigenstates are degenerate with respect to $ p$ and acquire the same structure as the usual Landau degeneracy. 
Nevertheless, the characteristics of the non-centrosymmetricity incorporated in the $ K_{ij} $-terms appears in the expression of $ T_c $. 
Similar to the previous case of $ D_4 $ with $ {\bf H} \parallel z $ the magneto-electric effect yields a reduction of the paramagnetic limiting term. This renormalization is surprisingly strong in general, as we can see when we return to the expressions which we had derived for the different coefficients. We obtain for the last term in Eq.(\ref{tc-c4v-hperp}), 
\begin{equation}
 \left [ \frac { K_{xy}^2}{4K_1}-Q_{xx} \right ]
\frac{H^2}{\alpha}=
\left [ \frac {\langle\phi^2({\bf k})
\hat{\bf{\gamma}}_x({\bf k})v_y({\bf k})\rangle^2}{\langle\phi^2({\bf k})v_x^2({\bf k})\rangle}-\langle\phi^2({\bf k})
\hat{\bf{\gamma}}_x^2({\bf k})\rangle \right ]\frac{\mu_B^2H^2N_+ S_3}{2\alpha}
\end{equation}
Considering the simplified picture of a parabolic band with $ {\bf v}({\bf k}) = {\bf k}/m^* $ and a Rashba spin-orbit coupling $ \hat{\bf{\gamma}}={\bf k}\times\hat z$ (setting $\gamma_z({\bf k})=0$) we find the amazing result that the two terms exactly cancel and the paramagnetic effect is completely suppressed. This effect can be immediately obtained, if we perform the gauge transformation 
\begin{equation}
{\bf q}\to{\bf q}+\frac{2\mu_B m^*(\hat z\times {\bf H})}{ k_F}.
\end{equation}
already in Eq.(\ref{Omega}) and so eliminating the paramagnetic term at the outset. However, it is important to notice that this exact cancellation is a consequence of the simplified forms of the band structure and the spin-orbit coupling term. Taking more realistic band structure effects into account it is obvious that this identity does not hold anymore in general. Nevertheless, our results suggests that
the magneto-electric effect can, in principle, yield a substantial contribution to eliminate the paramagnetic limiting also for fields in the basal plane.

\subsubsection{Symmetry $D_{4}$,  ${\bf H}\perp \hat z$}

It is easy to see that this case is analogue to the situation for the field along the $z$-axis and has only quantitative differences. 
Thus also here we encounter a reduction of the paramagnetic limit due to the magneto-electric effect yielding
\begin{equation}
T_c = T_{c0} -\frac{2e\sqrt{K_1K_2}}{\alpha}H + \left( \frac{K_{xx}^2}{3 K_1} - Q_{xx} \right) \frac{H^2}{\alpha} ,
\end{equation}
where also the same considerations concerning the gauge freedom apply as in the case of $ {\bf H} \parallel \hat{z} $ apply.

\subsubsection{Two-dimensional case, symmetry $C_{4v}$,  ${\bf H}\perp \hat z$}

The simplest way to pass  from 3D to 2D situation it is to introduce  $\delta (z)$ function potential well into 3D GL equation (\ref{3D}). It is equivalent to the theory used by Tinkham \cite{Tink} for the calculation of the upper critical field in a thin film with thickness $d<<\xi$ for a field parallel to the film.  Thus, we consider the instability equation 
\begin{eqnarray}
&\left\{\alpha(T-T_{c0})
- K_1(D_x^2+D_y^2)
-K_2\frac{\partial^2}{\partial z^2}+ K_{xy}(H_xD_y-H_yD_x) +Q_{xx}H^2\right.\nonumber\\
&\left.-\frac{2K_2}{d}\delta(z)
\right\}\eta=0,
\label{2D}
\end{eqnarray}
where $d$ is a length of the order of the film thickness that is in pure 2D case it  is an atomic scale length. This eigenvalue equation has the solution 

\begin{equation}
\eta({\bf r})=A\exp{\left [i({\bf p}\times {\bf r})_z\right ]}\exp{\left(-\frac{|z|}{d}\right)},
\end{equation}
where ${\bf p}=p{\bf H}/H$ is a vector with arbitrary length directed along magnetic field.  This then determines the critical temperature as a function of the applied magnetic field. 
\begin{equation}
\alpha(T-\tilde T_{c0})
+K_1(2eH)^2\langle(z-z_0)^2\rangle
 +\left (Q_{xx}-\frac{K_{xy}^2}{4K_1}\right )H^2
=0.
\end{equation}
Here $\tilde T_{c0}$ is the critical temperature in the absence of a magnetic field, corresponding to
$ d^2 = K_2 / \alpha(\tilde{T}_{c0}-T_{c0}) $. Moreover, 
$\langle...\rangle$ denotes the expectation value using  the wave function $\exp{(-|z|/d)}$ and 
 $z_0$ is determined by the same expression as in the 3D case
\begin{equation}
z_0=(2eH)^{-1}\left (p+\frac{K_{xy}}{2K_1}H  \right).
\end{equation}
Hence,  we obtain for the critical temperature
\begin{equation}
T_c=\tilde T_{c0}
+\left [ \frac { K_{xy}^2}{4K_1}-Q_{xx} \right ]
\frac{H^2}{\alpha}-\frac{K_1}{\alpha}(z_0^2+d^2/2)(2eH)^2.
\label{eq99}
\end{equation}
%Under some special conditions the terms in the square parenthesis cancel each other, as we had discussed above. 
The critical temperature reaches obviously a maximal value at $z_0=0$, i.e. for
\begin{equation}
p=-\frac{K_{xy}}{2K_1}H.
\end{equation}
%Thus, the degeneracy of solution in 2D case is not the case. Finally for the critical temperature we obtain
%\begin{equation}
%T_c=\tilde T_{c0}-
%\frac{K_1 d^2}{2\alpha}(2eH)^2.
%\label{para}
%\end{equation}
The upper critical field shows also here the square root temperature dependence usual for  thin films in a parallel magnetic field \cite{Tink}. Under special conditions (e.g. rotation symmetry around the normal vector of the film)  the expression in the square parenthesis in Eq. (\ref{eq99}) may vanish, as described above. Then, unlike in usual superconductors, non-centrosymmetric superconductors follow the standard Tinkham behavior unchanged by paramagnetic contributions. 

In view of strong inequality  $d\ll 1/\sqrt {2eH}$  the complete suppression of 2D superconducting state
($T_c(H)=0$) is reached in the field which exceeds
the orbital critical field in the 3D case (\ref{tc-c4v-hperp}).

\subsection{Two band case}

While the one-band picture discussed so far gives useful insights into the influence of the magneto-electric effect on the 
upper critical field, in particular, in the context of paramagnetic limiting, in reality there are at least two split bands whose
Fermi surface allows for the nucleation of a condensate in a finite magnetic field. In the two-band picture the situation is somewhat more complex, so that we restrict here to a few aspects only which, we believe, are relevant in this context
without attempting to give a complete overview. We base our analysis on the formalism introduced  for the homogeneous superconducting phase in section {\bf 3.2}. We use also an order parameter belonging to a one-dimensional representation on the two Fermi surfaces, labeled by $ \lambda = \pm $, $\tilde\Delta_\lambda({\bf k},{\bf r})=\eta_{\lambda}({\bf r})\phi_{\lambda}({\bf k})$. Moreover we restrict our discussion of the case of the point group $ C_{4v} $ with an in-plane magnetic field. Then the linearized Ginzburg-Landau equation is given by
\begin{eqnarray}
&\eta_{+}¥ =g_{++}[S_1(T)-\hat L_+]\eta_{+}+ 
g_{+-}[S_1(T)-\hat L_-]\eta_{-},\nonumber\\
&\eta_{-}=g_{-+}[S_1(T)-\hat L_+] \eta_{+}+
g_{--}[S_1(T)-\hat L_-]\eta_{-},
\label{e149}
\end{eqnarray}
with the operators $\hat L_\lambda$,
\begin{equation}
\hat L_\lambda=N_{0\lambda}^{-1}\left [K_{1\lambda}(D_x^2+D_y^2)
-K_{2\lambda}\frac{\partial^2}{\partial z^2}+\lambda K_{xy\lambda}(H_xD_y-H_yD_x) +Q_{xx\lambda}H^2\right ],
\end{equation}
where the coefficients are defined through the straightforward generalization of Eqs.(\ref{KK}) and (\ref{KQ}) to the two-band case with
the gap functions $\phi_{\lambda}({\bf k})$, the Fermi velocity 
components $v_{\lambda,i}({\bf k})$ and the densities of states $N_\lambda$ taken in the corresponding  band.

Similar to the one-band case the solutions of this equation system can be cast into the Abrikosov form
\begin{equation}
{\eta_+({\bf r})\choose \eta_-({\bf r})}={f_+(z)\choose f_-(z)}\exp{\left [i ({\bf p}\times {\bf r})_z\right ]}
\end{equation}
where again ${\bf p}=p{\bf H}/H$ and  the functions $f_+(z),~f_-(z)$ satisfy to the system of equations
\begin{eqnarray}
&f_+ =g_{++}[S_1(T)-\hat M_+]f_++ 
g_{+-}[S_1(T)-\hat M_-]f_-,\nonumber\\
&f_-=g_{-+}[S_1(T)-\hat M_+] f_++
g_{--}[S_1(T)-\hat M_-]f_-.
\label{e249}
\end{eqnarray}
Using the same gauge as in the one-band example, the new operator $\hat M_\lambda$ is then 
\begin{eqnarray} 
&\hat M_\lambda=N_{0\lambda}^{-1}\left [K_{1\lambda}(2eH)^2(z- z_{\lambda 0})^2
-K_{2\lambda}\frac{\partial^2}{\partial z^2} +\left (Q_{xx\lambda}-
\frac{K_{xy\lambda}^2}{4K_{1\lambda}}\right )H^2\right ],\\
&z_{0\lambda}=(2eH)^{-1}\left (p+\lambda\frac{K_{xy\lambda}}{2K_{1\lambda}}H  \right).
\end{eqnarray}
As in the one-band case the eigen states of this system possesses the Landau degeneracy represented through the
equilibrium positions of the coupled harmonic oscillators, $ z_{0+} $ and $ z_{0-} $, which both depend on $ p $. Through
the substitution
$$
z= Z+\frac{p}{2e H}
$$
we can formulate the equation system so that $ p $ is eliminated and $ z_{0\lambda} \to Z_{0\lambda} $,  
\begin{equation}
Z_{0\lambda}=\lambda\frac{K_{xy\lambda}}{4eK_{1\lambda}}.
\end{equation}

The general solution of Eq.(\ref{e249}) can be found only numerically. Here
we limit ourselves to a variational solution of the form,
\begin{eqnarray}
f_+(Z)=C_+\exp\left\{- \frac{eH\sqrt{K_{1+}}(Z-Z_{0+})^2}{\sqrt{K_{2+}}}\right\}, 
\nonumber\\
f_-(Z)=C_-\exp\left\{- \frac{eH\sqrt{K_{1-}}(Z-Z_{0-})^2}{\sqrt{K_{2-}}} \right\}.
\end{eqnarray}
In the following calculations, taking into account that the band splitting is much less than the Fermi energy  $|\gamma_\perp|k_F\ll\varepsilon_F$, we neglect the difference of the Fermi velocities 
$v_{\lambda}({\bf k})$ and the densities of states $N_{0\lambda}$ of the two bands. This case  the values of $Z_{0\lambda}=\lambda Z_0$ for different bands differ each other only by sign.
Returning to the free energy functional and integrating over $Z$ we obtain new variational equations for
the coefficients $ C_+ $ and $ C_- $:
\begin{eqnarray}
&C_+ =g_{++}[S_1(T)- M(H)]C_++ 
Ig_{+-}[S_1(T)-M(H)]C_-,\nonumber\\
&C_-=Ig_{-+}[S_1(T)-M(H)] C_++
g_{--}[S_1(T)-M(H)]C_-
\label{e259}
\end{eqnarray}
with 
\begin{equation}
{M} (H) =N_0^{-1}\left [
 2eH \sqrt{K_{1} K_{2}}
 +\left(Q_{xx} - \frac{K_{xy}^2}{4 K_{1}}\right)H^2 \right],
 \label{M}
\end{equation} 
and
\begin{equation}
I=\exp{(-2eHZ_0^2)}.
\end{equation}
The system (\ref{e259}) has the same form as the system (\ref{e49}) in the absence of magnetic field.
Hence for the critical temperature we obtain
\begin{equation}
T_c=\tilde T_{c0}(1-M)
\label{T_c}
\end{equation}
where he temperature $\tilde T_{c0}$ is given by the same formula 
(\ref{T_{c0}})
 as $T_{c0}$ taking in mind the substitutions $g_{+-}\to\tilde g_{+-}=Ig_{+-}$,~  $g_{-+}\to\tilde g_{-+}=Ig_{-+}$. The product $2eHZ_0^2\approx eH m^{*2}/(k_Fm)^2)$ is much less than unity generally, except for heavy fermion or layered superconductors. 
 
Thus, in the two-band situation 
the paramagnetic suppression of superconducting state  $\propto -Q_{xx}H^2$ is substantially weakened by the magneto-electric effect $\propto K_{xy}^2H^2/4K_1$. The latter has a finite value so long we work in the limit $\mu_BH\ll|\gamma_\perp| k_F$.

This conclusion is qualitatively valid in general, although our discussion was done by a simple variational approach only. Moreover, assuming $g_{++}=g_{--}$ and $g_{+-}=g_{-+}$, as it was done in the absence of magnetic field (see eqns. (\ref{s}) and (\ref{t})),  we come to the solution of (\ref{e259}) with either pure singlet or with pure triplet pairing. The effect of paramagnetic limiting takes place identically in both cases. This underlines directly that the weakening of the paramagnetic limiting in the non-centrosymmetric superconductors is not connected to the formation of a mixed singlet-triplet state. The important point is the spin-orbital splitting of the bands. The Pauli spin susceptibility of the quasiparticles  in whole space between the two Fermi surfaces is not changed in the superconducting state in comparison of its normal value that leads to the weakening of the paramagnetic suppression of the superconducting state.

%that neglecting the difference of the Fermi velocities and the density of states in two bands that is correct at  $|\gamma_\perp|k_F\ll\varepsilon_F$ we still keep the finite band splitting in comparison with the Zeeman energy $\mu_BH\ll|\gamma_\perp| k_F.$

Various approximations to the two-band model have been used in literature, some of which can obscure the subtleties of non-centrosymmetric superconductors. In particular, the discussion of the spin susceptibility in the superconducting phase as discussed in \cite{FAS04,Sam05,Sam07} have to be considered with caution in view of the magneto-electric effects which are neglected there. The adjustment of the superconducting state to the field-induced shifts of the Fermi surface yield a correction to the spin susceptibility which is not negligible as our discussion in the Ginzburg-Landau regime show. The subtle two-band effects, however, often make quantitative predictions difficult \cite{YS07}.

\section{Conclusion}

In this chapter we have given an overview on some theoretical aspects of non-centrosymmetric superconductors. Unlike symmetric spin-orbit coupling found in centrosymmetric metals, the antisymmetric spin-orbit coupling has a spectacular influence on the electronic bands through a specific spin splitting of the quasi-particle states. Superconductivity as a Fermi liquid instability is naturally influenced by such a modification of the electronic states. Superconductivity then has always a multi-band character. Moreover, parity does no longer provide a good quantum number to classify the superconducting phases. 

One of the physically most remarkable aspects of non-centrosymmetric superconductivity is connected with magneto-electricity, the peculiar connection between supercurrents and spin polarization. 
We have considered one aspect in this context, namely its influence on paramagnetic limiting. This effect is of interest in strongly correlated electron systems where the coherence length is generally small due to
the enhanced masses like in heavy Fermion compounds. Here ordinary orbital depairing in a magnetic field is weak, such that the upper critical field reaches magnitudes where paramagnetic limiting through spin polarization becomes visible. Interestingly already on the basis of symmetry considerations it is possible to arrive at very interesting predictions which are borne out in some of the non-centrosymmetric heavy Fermion superconductors.


\begin{thebibliography}{99.}
\bibitem{footnote}The interband pairing leads to the interesting possibility of existance of nonuniform superconducting states even in the absence of external magnetic fields. \cite{MinSam08}  These states, however, could realized in the noncentrosymmetric compounds with the SO band splitting smaller than the superconducting critical temperatue. To the best of our knowledge, in all noncentrosymmetric compounds discovered to date the relation between the two energy scales is exactly the opposite: the SO band splitting exceeds all superconducting energy scales by order of magnitude, completely suppressing the interband pairing, both uniform and nonuniform.
\end{thebibliography}
\end{document}